\documentstyle[emulateapj]{article}

\slugcomment{The Astrophysical Journal (Letters), in press}

\lefthead{Ouchi et al.}
\righthead{Sub-mm \& dust properties of LBGs}

\begin{document}

\title{Expected Sub-mm Emission and Dust Properties \\
of Lyman Break Galaxies at High Redshift}

\author{Masami Ouchi \altaffilmark{1}, Toru Yamada \altaffilmark{1}}
\and
\author{Hideaki Kawai \altaffilmark{2}, Kouji Ohta \altaffilmark{2}}

\altaffiltext{1}{Astronomical Institute, Tohoku University, Aoba-ku, 
Sendai 980-8578, Japan\\
e-mail: ouchi/yamada@astr.tohoku.ac.jp } 
\altaffiltext{2}{Department of Astronomy, Faculty of Science, Kyoto University, Sakyo-ku, 
Kyoto 606-8502, Japan \\
e-mail: ykawai/ohta@kusastro.kyoto-u.ac.jp }

\begin{abstract}
 We investigate expected sub-mm emission and dust properties of the Lyman Break Galaxies (LBGs) in the Hubble Deep Field (HDF). The SCUBA Deep Survey (\cite{hug98}) provides an upper limit of the $850 \mu$m flux densities of the LBGs. With this constraint, we argue that a typical ultra-violet to far-infrared spectral shape of the high-redshift LBGs is rather close to a template spectrum of low-reddening starburst galaxies in the local universe but different from that of heavily dust-enshrouded ultra-luminous far-infrared galaxies like Arp220.
We also evaluate the lower-limit temperature of dust in LBGs assuming single- and two-component modified black-body spectrum. To estimate the total amount of energy re-emitted in FIR wavelength, we take two different approaches; model fitting of the UV spectra of LBGs and an empirical method using the relationship between UV spectral shape and UV/FIR flux ratio observed for local starburst galaxies. Both methods give lower-limit temperature of the LBGs as $\sim 40$K, which is higher than the typical dust temperature of local optical- and infrared-selected galaxies. This result is also supported by the comparison of the expected sub-mm flux of the LBGs with the cosmic FIR background radiation. The high dust temperature may indicate the effective massive-star formation or different dust properties in the high-redshift LBGs.
\end{abstract}


\section{Introduction}

 A number of high-redshift star-forming galaxies have been identified through the Lyman break surveys (Steidel et al. 1996a, b; Lowenthal et al. 1997). Since the number density of these Lyman Break Galaxies (LBG) is comparable to or somewhat higher than that of the local $\sim L_*$ galaxies, they are thought to be natural candidates of the progenitors of nearby normal galaxies (Lowenthal et al. 1997).
 From the observed UV continuum emission, a typical star-formation rate (SFR) of LBGs is estimated to be moderate, $\sim 10$ M$_\odot$ yr$^{-1}$ ($H_0$=50 km s$^{-1}$ Mpc$^{-1}$ and $q_0$=0.5).  However, there are increasing evidences of the presence of significant amount of extinction in the UV continuum by interstellar dust in LBGs (\cite{petprs}; \cite{saw98}).  If the inferred UV extinction is corrected, their SFR would be more than $\gtrsim 100 M_\odot$ yr$^{-1}$ (Dickinson 1998), which may imply that high-redshift LBGs are in their major star-formation period.

 On the other hand, deep surveys in sub-mm wavelength with Sub-millimeter Common-User Bolometer Array (SCUBA; Holland et al. 1998) revealed a population of FIR-luminous galaxies possibly at high redshift and have a large SFR, $\gtrsim 100$ M$_\odot$ yr$^{-1}$ (Smail et al 1997; Hughes et al. 1998; Barger et al. 1998; Lilly et al. 1999). Although the optical identification of the SCUBA sources has not been well established in some cases, they are also candidates of high-redshift forming galaxies. 

 It is then natural to ask what the relationship between these two populations is, or, how the high-redshift LBGs appear in sub-mm wavelength. In the deepest sub-mm imaging so far at the Hubble Deep Field (HDF; Williams et al. 1996) by Hughes et al. (1998), none of the optically-selected LBGs (Steidel et al. 1996b; 
Lowenthal et al. 1997) was detected. The upper-limit flux value at 850 $\mu$m, however, can be used to put some important constraints for the spectral energy distribution (SED) of the LBGs. In this Letter, we investigate sub-mm emission properties of the LBGs in HDF to constrain their physical properties such as dust temperature by using the results of the SCUBA observations by Hughes et al. (1998). 

\section{Spectral Shape of the Lyman Break Galaxies}

Our sample of LBGs is the same as that used in Sawicki and Yee(1998). They are the 17 galaxies at z$\sim 3$ with firm spectroscopic redshifts and with photometric data from optical to near-infrared (NIR) wavelength, taken from the 41 high-redshift LBG candidates originally provided by Steidel et al. (1996b) and Lowenthal et al. (1997). Redshifts and magnitudes are listed in Table 1 of Sawicki and Yee (1998).

 Hughes et al. (1998) detected five sources at 850 $\mu$m with SCUBA
in HDF. We found none of the optically-selected LBGs is detected within the error circles of the SCUBA sources \footnote{ There may be small systematic errors in the positions quoted in Hughes et al. (Richards 1998) but it does not affect the identification with LBGs discussed here.  }.
 Hughes et al. evaluated their detection limit  $\sim 2$ mJy (5$\sigma$). Hereafter we quote this value as the upper-limit flux density at 850 $\mu$m for all the LBGs although the true detection limit may change across the field slightly due to the position-dependent confusion noises and detector sensitivities.  

\begin{figure}
\plottwo{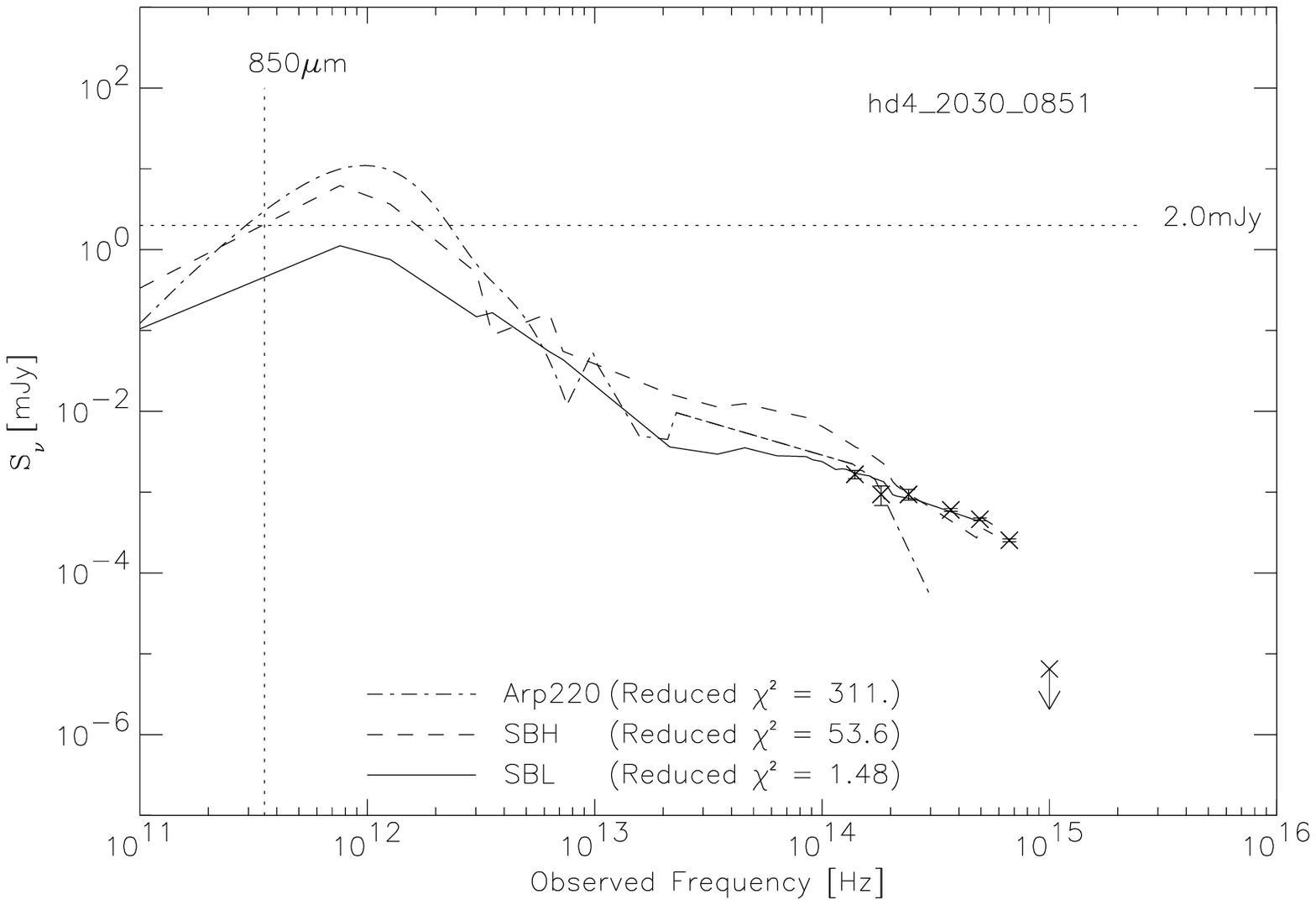}{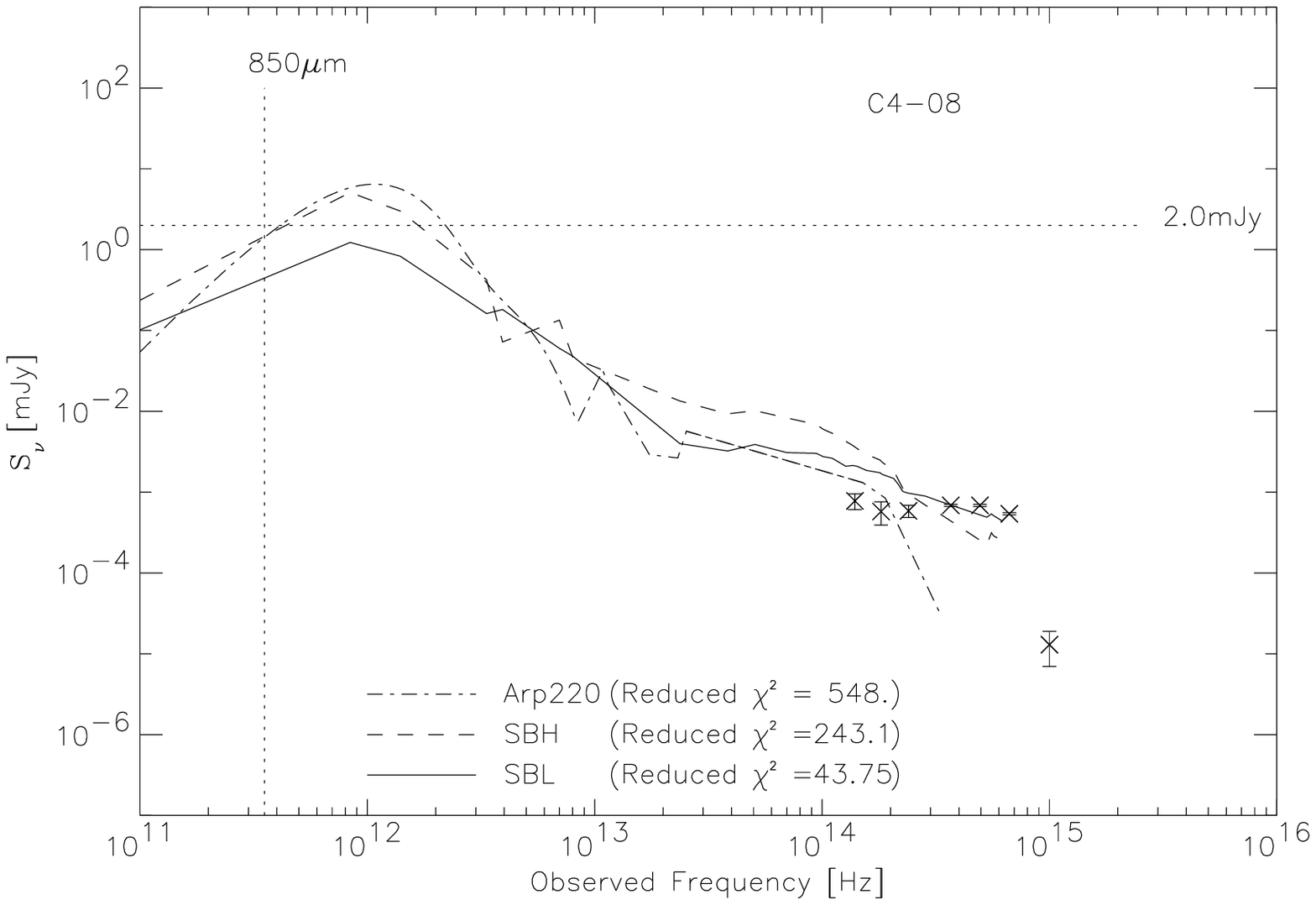}
\caption{ SED of the LBGs is fitted by the templates of local starburst galaxies; high-reddening starburst (dashed line) and low-reddening  starburst (solid line) compiled by Schmitt et al. (1997) and Arp220 (Klaas et al. 1997; dash-dotted line). Examples of the best and the worst fittings are shown in panel (a) and (b), respectively. The $U$-and $B$-band data are not used in the fitting.  
\label{fig1}} 
\end{figure}

 The upper limit of sub-mm flux density gives a constraint on the spectral shape from the rest-frame UV to FIR wavelength of the LBGs. We compare the observed optical-NIR flux densities and the upper limit of sub-mm flux density of LBGs with the three template spectra of local starburst galaxies. For the template spectra, we adopt the SED of Arp220 (Klaas et al. 1997), as an example of heavily dust enshrouded starburst galaxies, and the averaged SED of high-reddening starburst galaxies (SBH) and low-reddening starburst galaxies (SBL) compiled by Schmitt et al. (1997).
 These templates are redshifted and then fitted to the optical-NIR
photometric data. Figure~\ref{fig1}a and \ref{fig1}b show the best and
the worst case in the results of our least-square fitting
procedure. In many cases, the template of SBL gives a better fit but
those of Arp220 and SBH are always steeper than the spectra of LBGs
(Figure~\ref{fig1}a). For several objects, spectra of LBGs are even flatter than the SBL template (Figure \ref{fig1}b). Reduced $\chi ^2$ values (degree of freedom is four) ranges 300-550 for Arp220, 50-250 for SBH, and 1.5-30 for SBL. The absorption at the rest-frame UV wavelength in the LBGs thus seems much smaller than in Arp220- or SBH-like galaxies.  
 If we assume the templates of Arp220 and SBH (fitted at optical-NIR wavelength), the expected flux density at $850 \mu$m exceeds 2 mJy for 11 and 9 of the 17 objects, respectively. If the SBL template is applied, none of the objects has a $850 \mu$m flux density larger than 2 mJy, which is consistent with the results of the SCUBA observation. Thus, not only the rest-frame UV-optical SEDs but also the FIR-UV SEDs of LBGs are better represented by the SBL template.  

 On the other hand, SCUBA-selected bright sub-mm sources show more FIR excess. For example, Barger et al. (1998) detected a sub-mm source with $S_{\nu (850 \mu m)}=4.6$ mJy by SCUBA in the Lockman-Hole field and found a plausible optical counterpart with $K_{AB} = 21.8$. They argued that only the SED of ultra-luminous infrared galaxies like Arp220 can match this flux ratio. The SEDs of the LBGs are thus much shallower than those of the SCUBA sources detected above a few mJy level at 850 $\mu$m.  The observed star-forming regions of LBGs are not likely to be heavily enshrouded by dust compared with the SCUBA-selected bright sub-mm sources or the local ultra-luminous infrared galaxies.  

\section{Dust Properties of Lyman Break Galaxies} 

\subsection{Model Fitting of the UV Spectra} 

 We first estimate the amount of dust extinction of LBGs by fitting
the model UV/optical spectra (at 900-10000 \AA ; Salpeter IMF is
assumed ) with an extinction to the observed SEDs following the
similar manner as in Sawicki and Yee (1998) but using the different
evolutionary-synthesis model developed by Kodama \& Arimoto (1997).
We assumed the reddening law derived by Calzetti for the local
starburst galaxies \footnote{The formula is presented in Sawicki and
Yee (1998)}. The estimated extinction, $E(B-V)$, ranges 0.3-0.5 mag
with ages of $\sim $ 0.03 Gyr, which agrees with the results in
Sawicki \& Yee (1998). For 13 of the 17 objects, acceptable fittings
are obtained with finite amounts of extinction and we will discuss the
13 objects in this section. The $850\mu$m flux densities of LBGs are
then calculated, with the assumption that all the energy absorbed in
UV/optical wavelength is re-emitted by the modified black-body
radiation with single temperature. The observed 850 $\mu$m flux
density, $S_{\nu (850 \mu m)}$, is given by  

\begin{equation}
S_{\nu (850 \mu m)}=
\frac{(1+z) F_{abs}}{\int_{0}^{\infty} {\nu}^{\beta} B_{\nu}(T) d\nu} 
{\nu_1}^{\beta} B_{\nu_1}(T)
\end{equation}

where $F_{abs}$ is absorbed flux, $T$ is a dust temperature, $\beta$
is a power of the emissivity law, $B_{\nu}(T)$ is the Planck function,
and $\nu_1=(1+z) \nu_{850 \mu m}$.   
 
\begin{figure}
\plotone{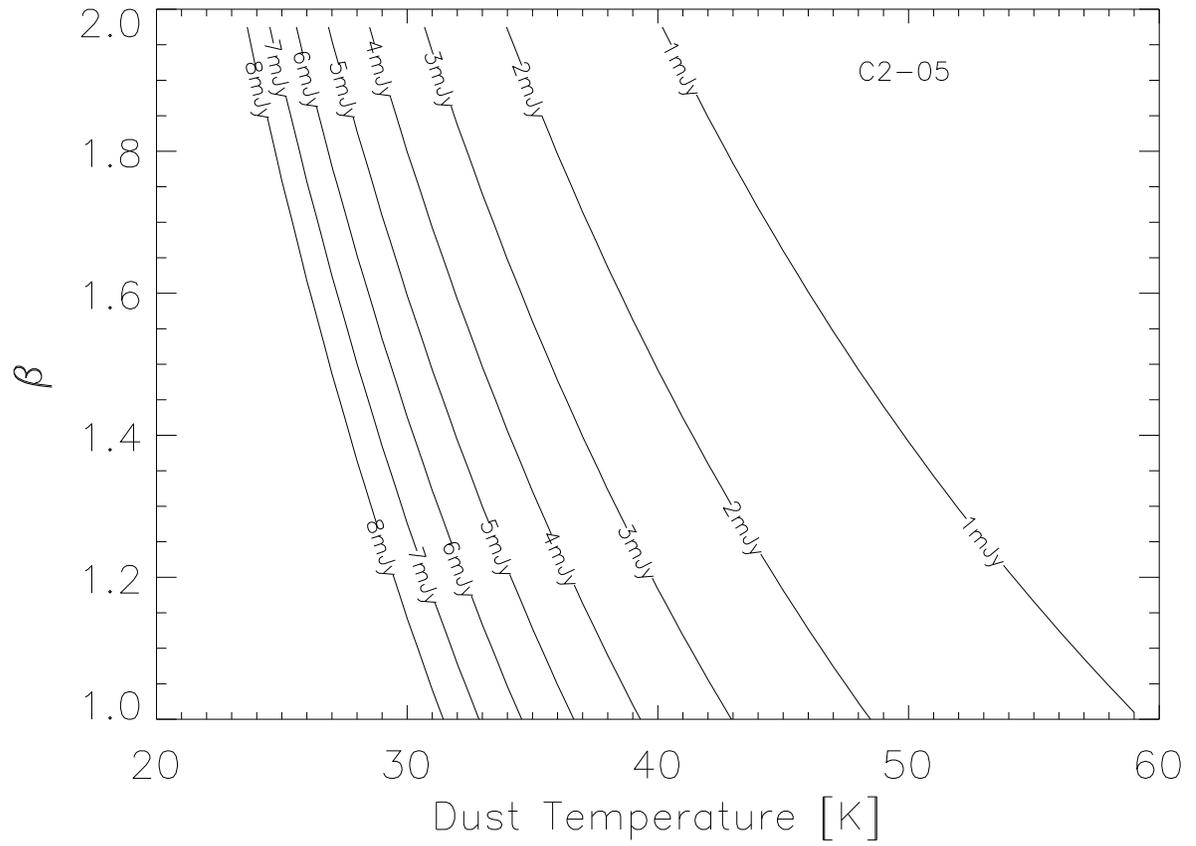}
\caption{An example of the dependence of the estimated flux
density at 850 $\mu$m on the dust temperature and the power of dust
emissivity law $\beta$. 
\label{fig2}}
\end{figure}

\begin{figure}
\plotone{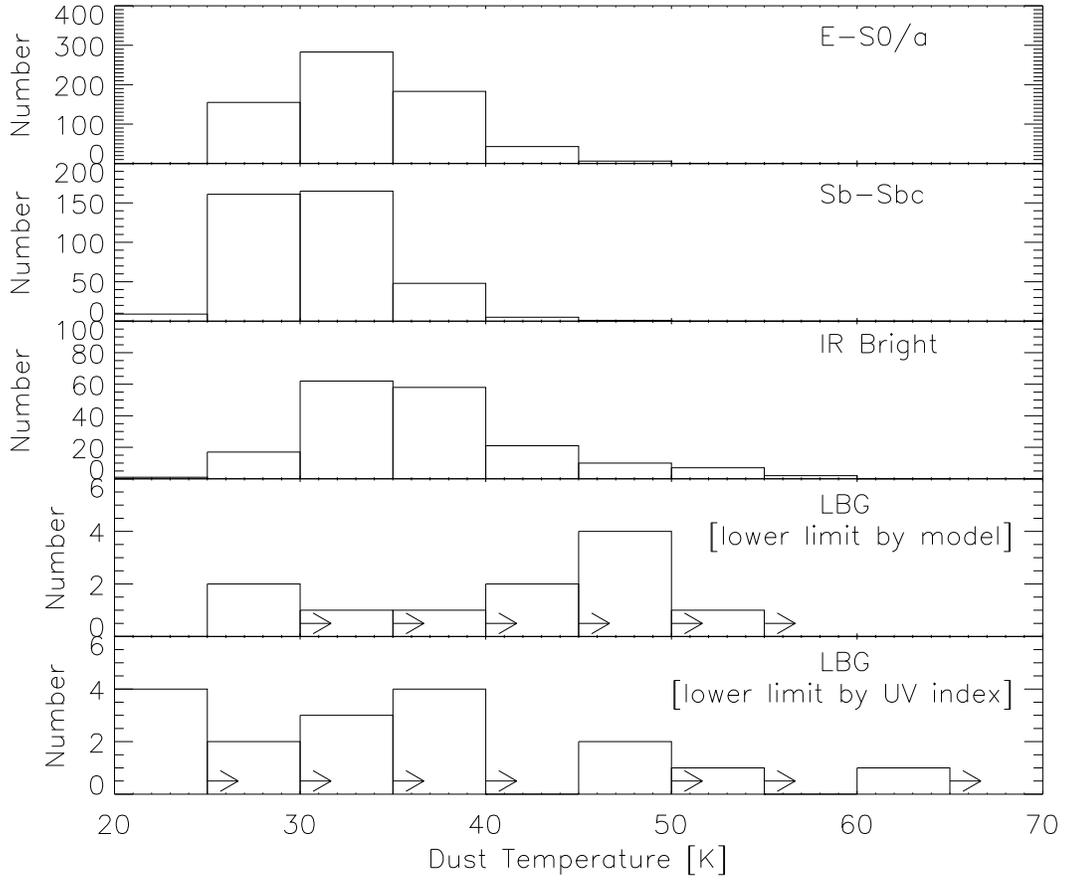}
\caption{ The bottom two panels show the distributions of
the lower-limit dust temperature of LBGs evaluated by modeling the UV
extinction (single dust temperature assumed) and by using the
empirical method (see text).  The top three panels show the dust
temperature of the local optically- and infrared-selected galaxies in
literatures.\label{fig3}} 
\end{figure}

 Figure \ref{fig2} shows the dependence of the estimated flux density
on the dust temperature and emissivity for a typical case. We assume 
$\beta =1$ in the following discussion, in order to compare with the 
properties of the local galaxies which are obtained with $\beta =1$.
 For the given $\beta$ value, the upper limit of the flux density at
850 $\mu$m (2  mJy) imposes the lower limit of the dust temperature,
$T_{lim}$. In Figure \ref{fig3} (the fourth panel from the top), the
distribution of the obtained $T_{lim}$ is shown along with those of
the dust temperature of local galaxies from the samples of the
optically-selected CfA galaxies classified as E-S0/a and Sb-Sbc
(Sauvage and Thuan 1994)  and the $IRAS$-selected infrared-bright
galaxies (Young et al. 1989). These dust temperature of the local
galaxies are measured by fitting the $IRAS$ 60 and 100 $\mu$m flux
densities with single-temperature modified-Planck function with
$\beta=1$.
 The distribution of $T_{lim}$ of the LBGs has a peak at around 45 K
and significantly higher than those of the local galaxies. The median
value of the temperature {\it lower limit} is 43 K while those of CfA
and infrared-bright samples are $T_{med}= 33$ K (E/S0), 30 K (Sb-Sbc),
and 36 K (infrared bright). According to the Peto-Prentice Generalized
Wilcoxon Test, the temperature distribution of LBGs is different from
those of the local galaxies with a confidence level larger than
99.9999 \% .

 It is possible that the high dust temperature of the LBGs is
introduced by our assumption of single dust temperature. Most of the
nearby star-forming galaxies have a hot dust component ($T \sim
100-150$ K) in addition to a cold one ($T \sim 30-50$ K) (Eales,
Wynn-Williams, and Duncan 1989; Klaas et al. 1997). IRAS 60 and 100
$\mu$m flux densities of the local galaxies may be dominated by only
the cold component. So as to estimate the energy drained to the hot
dust, we fitted  the SEDs  of the three template spectra used in the
previous section by the two component models. The results are
$L(cold)/L(hot)=1.9$ for the SBL, $L(cold)/L(hot)=3.2$ for the SBH,
and $L(cold)/L(hot)=9.8$ for the Arp220 template (The result for
Arp220 is consistent with that of Klaas et al.). 
 If we consider the contribution by the possible hot component in
calculating the FIR luminosity of the LBGs, adopting the ratio
$L(cold)/L(hot)=1.9$ which gives the lowest limiting temperature, the
peak of $T_{lim}$ of the LBGs is shifted to $\sim 37$ K.  According to
the Peto-Prentice Generalized Wilcoxon Test, they are still different
higher than 99.9999 \% level. True $T_{lim}$ must be between these two
extreme cases, and it may be $T_{lim} \sim 40$ K. In any case, the
dust temperature of the LBGs are likely to be still higher than those
of the nearby normal galaxies. 

 The difference of local galaxies and high-redshift LBGs can be in
dust extinction properties instead of dust temperature. The reddening
law in the LBGs is not understood very well (e.g., Pettini et
al. 1998). If we use the SMC-like reddening law instead of the
Calzetti law, the total amount of absorbed UV luminosity becomes
smaller and the expected sub-mm flux density becomes several times
smaller. In such a case, only a weak constraint for dust temperature
is obtained. 

\subsection{FIR emission Estimated from UV spectral Slope}

 UV spectra of the local starburst galaxies are well approximated by a
power-law of index $p$ (1250 $\sim 2200$ \AA ), $f_\lambda=A
\lambda^p$ (\cite{cal94}, \cite{meu95}), and there is an empirical
relation between $p$ and $F_{FIR}/F_{UV}$ values (\cite{meu97}). In
the range of $p \gtrsim -1.8$, approximately, 

\begin{equation} \label{eq of b-fir}
R \equiv \log \left ( \frac{F_{FIR}}{F_{UV}} \right ) \approx 0.65 p +1.8
\end{equation}

where $F_{FIR} = 1.26 \times 10^{-11} \left [ 2.58 \left
( \frac{S_{\nu(60 \mu m)}}{{\rm Jy}} \right )+\left( \frac{S_{\nu (100 \mu
m)}}{{\rm Jy}} \right) \right ] $  \ ${\rm erg}\, {\rm s}^{-1}$ ${\rm
cm}^{-2}$, and $F_{UV} = \lambda_c f_{\lambda_c} $ ($\lambda_c=2320$
\AA ). Hence, 

\begin{equation}
F_{FIR} \approx A {\lambda_c}^{p +1} \times 10^R.
\end{equation}

 If we assume the modified black-body radiation with the power of
emissivity law $\beta=1.0$, the dust temperature $T$ implies the flux
ratio $k(T) \equiv S_{\nu (100 \mu m)}/S_{\nu (60 \mu m)}$. Then  we
obtain,  

\begin{equation} \label{eq of f60}
S_{\nu (60 \mu m)}=\frac{1}{1.26 \times 10^{-11}(2.58+k(T))}
\left( \frac{F_{FIR}}{{\rm erg}\, {\rm s}^{-1}\, {\rm cm}^{-2}}\right) \ \ {\rm Jy}.
\end{equation}

 The sub-mm flux is predicted by the spectrum from eq.\ref{eq of
f60}. Again, the upper limit of $ S_{\nu \, lim (850 \mu m)} = 2$ mJy
gives  $T_{lim}$ values. The bottom panel of Figure \ref{fig3} shows a
distribution of $T_{lim}$ obtained by this empirical method.  It peaks
at $\sim38 K$, which is similar to or somewhat lower than the value
calculated from the model UV extinction. It is not surprising that the
empirical method gives wider $T_{lim}$ distribution since there is
fairly large scatter in the empirical relation (eq.\ref{eq of b-fir})
itself (Meurer et al 1997). 

\section{Contribution to the CFIRB}

 The average dust temperature of LBGs can also be examined with the observed intensity of cosmic far-infrared background (CFIRB) at $850 \mu$m, although the field of view of HDF (4.7 arcmin$^2$) is very small and statistical uncertainty may not be negligible; a typical fluctuation of CFIRB at this scale is still unknown.
 The total $850 \mu$m flux in HDF evaluated from the average intensity of the CFIRB radiation, $\nu I_{\nu} = 3.5 \times 10^{-10}$ W m$^{-2}$ sr$^{-1}$ (Fixsen et al. 1997). The five SCUBA sources contribute about 45 \% of this flux. The contribution of the LBGs in HDF should not significantly exceed the remaining 55\% of the CFIRB flux.  

\begin{figure}
\plotone{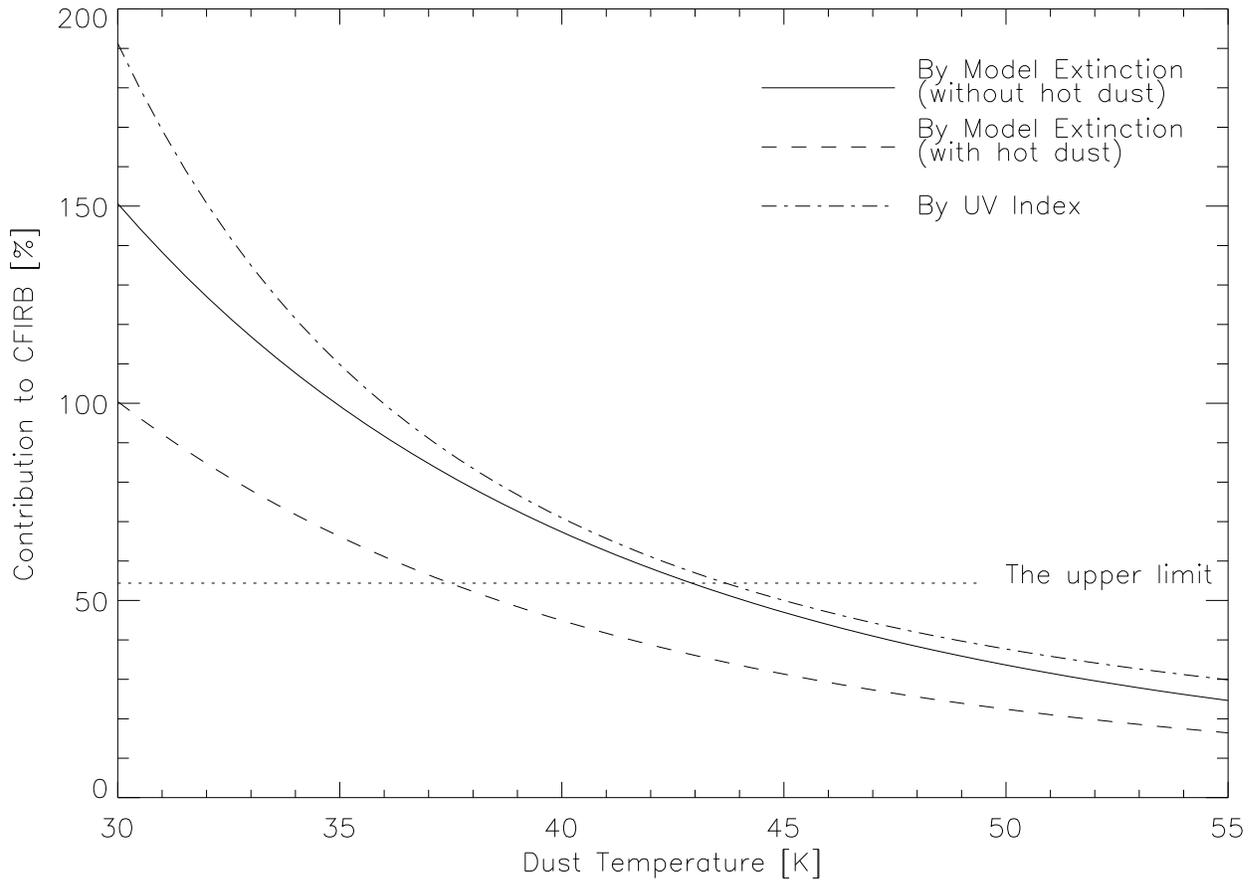}
\caption{The expected contribution of LBGs to the CFIRB
at $850 \mu$m is plotted in function of the assumed dust
temperature. The cases of the model extinction with single- and
two-component dust emission (solid and dashed line, respectively) and
the case of the empirical method (dash-dotted line) are shown.  
\label{fig4}} 
\end{figure}

  In Figure \ref{fig4}, we plot the contribution of the LBGs studied in the previous section to the CFIRB with respect to the assumed dust temperature. Three curves are those given by the model calculations with single and double temperature components and the prediction using the empirical relation.  Depending on  temperature, LBGs contribute a significant fraction ($\gtrsim 30\%$). With the constraint not to exceed the CFIRB flux, we obtain fairly high limiting temperature, $T_{lim} \sim 40$ K, which is consistent with the limit obtained by using the SCUBA upper-limit flux density.

\section{Discussions}

 We showed that the LBGs in HDF have UV-to-FIR SEDs different from
those of the ultra-luminous FIR galaxies in the local universe and of
the SCUBA-selected bright sub-mm sources. It may imply that the birght
sub-mm sources detected above a few mJy at 850 $\mu$m are rather
extremely dust-rich objectes among the high-redshift star-forming
galaxies selected by the Lyman break technique although it does not have to mean that the entire parent populations of sub-mm sources and LBGs are fundamentally different.

 We also claimed that dust temperature in the LBGs should be higher than $\sim 40$K, if we assume the reddening law appropriate for local starburst galaxies. What causes the relatively high dust temperature in the LBGs? High dust temperature may be results of some intrinsic properties of star formation in high-redshift galaxies. There are some possible explanations for effective dust heating.  For example,  a flatter initial mass function (IMF) which produces a larger number of massive stars for a given mass of dust may provide the effective dust heating. High star-formation efficiency (a higher SFR for a given mass of dust regardless of IMF) brings the similar situation. The difference may also be in properties of dust grains. For the case of low metal abundance (possibly true in young galaxies), dust may be dominated by small-size grains which are easily heated.

 Finally, we discuss the detectability of LBGs in sub-mm observations. Assuming an appropriate spectral shape like SBL or dust temperature of $40-50K$, we predict that sub-mm flux at $850\mu$m of the LBGs in HDF is typically 0.1-1 mJy.  So within the reasonable amount of observational time, only the brightest LBGs could be detected with SCUBA. In the coming decade, however, new sub-mm observational facilities such as Large Millimeter and Sub-millimeter Array (NAOJ) or MilliMeter Array (NRAO), etc. are planned. Our results presented in this letter provides rather optimistic future prospects; only a few hour exposure time will be needed to detect typical LBGs with such new facilities and then we will learn about true star-formation properties of typical high-redshift galaxies. 

\acknowledgments
 We thank N.Arimoto and T. Kodama for kindly providing their evolutionary synthesis code. 


\clearpage

\clearpage

\begin{thebibliography}{}

\bibitem[Barger et al., \ 1998]{bar98} Barger, A. J. et al. 1998, Nature, 394, 248

\bibitem[Calzetti Kinney \& Storchi-Bergmann 1994]{cal94} 
Calzetti, D., Kinney, A. L. \& Storchi-Bergmann, T. 1994, \apj, 429, 582 

\bibitem[Dickinson 1998]{dic98} 
Dickinson, M. 1998, in The Hubble Deep Field, eds. M. Livio, S.M. Fall and P. Madau. in press (astro-ph/9802064)

\bibitem[Eales, Williams, \& Duncan, \ 1989]{eal89} Eales, S. A.,
Wynn-Williams, C. G., \& Duncan, W. D. 1989, ApJ, 339, 859

\bibitem[Fixsen et al. 1997]{1997ApJ...490..482F} Fixsen, D. J., et al. 
1997, \apj, 490, 482 

\bibitem[Heckman et al., \ 1998]{hec98} Heckman, T. M., Robert, C.,
Leitherer, C., Barnett, D. R., \& Rydt, F., 1998, ApJ, 503, 646

\bibitem[Holland et al. 1998]{holla98} Holland, W., et al., 1998, \mnras, in press (astro-ph/9809122)

\bibitem[Hughes et al., \ 1998]{hug98} Hughes, D. H., et al. 1998, Nature, 394, 241

\bibitem[Helou et al.,\ 1988]{hel88} Helou, G., Khan, I. R., Malek, L., \&
Boehmer, L. 1988, ApJS, 68, 151

\bibitem[Klaas et al.,\ 1997]{kla97} Klaas, U., Martin, H., Heinrichsen, I., \& Schulz, B. 1997,
A\&A, 325,L21

\bibitem[Kodama \& Arimoto 1997]{1997A&A...320...41K} Kodama, T. \& 
Arimoto, N. 1997, \aap, 320, 41 

\bibitem[Leitherer \& Heckman \ 1995]{lei95} Leitherer, C., \& Heckman,
T. 1995, ApJS, 96, 9 

\bibitem[Lilly et al. 1999]{lilly99}  Lilly, S. J. et al. 1999, \apj, in press (astro-ph/9901047)

\bibitem[Lowenthal et al.,\ 1997]{low97} Lowenthal, J. D., Koo, D. C., Guzman, R.
, Gallego, J., Phillips, A. C., Vogt, N. P., Illingworth, G. D., \& Gronwall, C. 1997, ApJ, 481,673 

\bibitem[Meurer et al., \ 1995]{meu95} Meurer, G., Heckman, T.,
Leitherer, C., Kinney, A., Robert, C., \& Garnett, D. 1995, AJ, 110,
2665

\bibitem[Meurer et al.,\ 1997]{meu97} Meurer, G., Heckman, T.,
Leitherer, C., Lowenthal, J., \& Lehnert, M. 1997, AJ, 114, 54 

\bibitem[Pettini et al. 1998]{petprs} Pettini, M., Kellogg, 
M., Steidel, C. C., Dickinson, M., Adelberger, K. L. \& Giavalisco, M. 
1998, \apj, 508, 539


\bibitem[Richards,\ preprint]{ricprp} Richards, E. A. 1998, astro-ph/9811098

\bibitem[Sauvage \& Thuan 1994]{sau94} Sauvage, M. \& Thuan, 
T. X. 1994, \apj, 429, 153 

\bibitem[Sawicki \& Yee \ 1998]{saw98} Sawick, M., \& Yee,
H. K. C. 1998, AJ, 115, 1329

\bibitem[Schmitt et al.,\ 1997]{sch97} Schmitt, H. R., Kinney, A. L.,Calzetti, D., \& Bergmann, T. S. 1997, AJ, 114, 592S


\bibitem[Smail Ivison \& Blain 1997]{1997ApJ...490L...5S} Smail, I., Ivison, R. J. \& Blain, A. W. 1997, \apjl, 490, L5 

\bibitem[Steidel et al.,\ 1996a]{ste96a} Steidel, C. C., Giavalisco, M., Dickinson, M., \& Adelberger, K. L. 1996a, AJ, 112, 352

\bibitem[Steidel et al.,\ 1996b]{ste96b} Steidel, C. C., Giavalisco, M., Pettini, 
M., Dickinson, M., \& Adelberger, K. L. 1996b, ApJ, 462, L17

\bibitem[Steidel et al.,\ preprint]{steprp} Steidel, C. C.,
Giavalisco, M., Pettini, M., Dickinson, M., \& Adelberger, K. L. 1998, 
preprint.(astro-ph/9811399)

\bibitem[Williams et al. 1996]{1996AJ....112.1335W} Williams, R. E., et al. 
1996, \aj, 112, 1335 

\bibitem[Young Xie Kenney \& Rice 1989]{you89} Young, J. S., 
Xie, S., Kenney, J. D. P. \& Rice, W. L. 1989, \apjs, 70, 699 

\end{thebibliography}
\end{document}